%% file: conference_101719.tex
\documentclass[conference]{IEEEtran}
\IEEEoverridecommandlockouts
\usepackage{cite}
\usepackage{amsmath,amssymb,amsfonts}
\usepackage{algorithmic}
\usepackage{graphicx}
\usepackage{textcomp}
\usepackage{xcolor}
\usepackage[nolist]{acronym}
\usepackage{paralist}
\usepackage{booktabs}
\def\BibTeX{{\rm B\kern-.05em{\sc i\kern-.025em b}\kern-.08em
    T\kern-.1667em\lower.7ex\hbox{E}\kern-.125emX}}

\begin{document}


\input{acronyms}

\bstctlcite{IEEEexample:BSTcontrol}

\title{Investigating the Cybersecurity of Smart Grids Based on Cyber-Physical Twin Approach}

\author{
\IEEEauthorblockN{%
Ömer Sen\IEEEauthorrefmark{1},
Florian Schmidtke\IEEEauthorrefmark{1},
Federico Carere\IEEEauthorrefmark{2},
Francesca Santori\IEEEauthorrefmark{2},
Andreas Ulbig\IEEEauthorrefmark{1},
Antonello Monti\IEEEauthorrefmark{3}
}

\IEEEauthorblockA{%
\IEEEauthorrefmark{1}\textit{IAEW, RWTH Aachen Univserity,} Aachen, Germany\\
Email: \{o.sen, f.schmitdke, a.ulbig\}@iaew.rwth-aachen.de}
\IEEEauthorblockA{%
\IEEEauthorrefmark{2}\textit{ASM Terni S.p.A.} Terni, Italy\\
Email: \{federico.carere, francesca.santori\}@asmterni.it}
\IEEEauthorblockA{%
\IEEEauthorrefmark{3}\textit{ACS, RWTH Aachen University,} Aachen, Germany\\
Email: amonti@eonerc.rwth-aachen.de}
}

\IEEEoverridecommandlockouts

\IEEEpubid{\makebox[\columnwidth]{978-1-6654-3254-2/22/\$31.00~\copyright2022 IEEE \hfill}\hspace{\columnsep}\makebox[\columnwidth]{ }}

\maketitle

\IEEEpubidadjcol

\begin{abstract}
While the increasing penetration of information and communication technology into distribution grid brings numerous benefits, it also opens up a new threat landscape, particularly through cyberattacks.
To provide a basis for countermeasures against such threats, this paper addresses the investigation of the impact and manifestations of cyberattacks on smart grids by replicating the power grid in a secure, isolated, and controlled laboratory environment as a cyber-physical twin.
Currently, detecting intrusions by unauthorized third parties into the central monitoring and control system of grid operators, especially attacks within the grid perimeter, is a major challenge.
The development and validation of methods to detect and prevent coordinated and timed attacks on electric power systems depends not only on the availability and quality of data from such attack scenarios, but also on suitable realistic investigation environments.
However, to create a comprehensive investigation environment, a realistic representation of the study object is required to thoroughly investigate critical cyberattacks on grid operations and evaluate their impact on the power grid using real data.
In this paper, we demonstrate our cyber-physical twin approach using a microgrid in the context of a cyberattack case study.
\end{abstract}


\input{chapter1}
\input{chapter2}
\input{chapter3}
\input{chapter4}
\input{chapter5}

\noindent\textsc{Acknowledgments}\hspace{1em}
This work has partly been funded by the European Union’s Horizon 2020 research and Innovation programme under grant agreement N°832989.

\bibliographystyle{IEEEtran}
\bibliography{conference_101719}

\end{document}

%% file: acronyms.tex
\begin{acronym}
\acro{sg}[SG]{smart grid}
\acroplural{sg}[SGs]{smart grids}
\acro{der}[DER]{distributed energy resource}
\acroplural{der}[DERs]{distributed energy resources}
\acro{ict}[ICT]{information and communication technology}
\acro{fdi}[FDI]{false data injection}
\acro{scada}[SCADA]{Supervisory Control and Data Acquisition}
\acro{mtu}[MTU]{Master Terminal Unit}
\acroplural{mtu}[MTUs]{Master Terminal Units}
\acro{hmi}[HMI]{Human Machine Interface}
\acro{plc}[PLC]{Programmable Logic Controller}
\acroplural{plc}[PLCs]{Programmable Logic Controllers}
\acro{ied}[IED]{Intelligent Electronic Device}
\acroplural{ied}[IEDs]{Intelligent Electronic Devices}
\acro{rtu}[RTU]{Remote Terminal Unit}
\acroplural{rtu}[RTUs]{Remote Terminal Units}
\acro{iec104}[IEC-104]{IEC 60870-5-104}
\acro{apdu}[APDU]{Application Protocol Data Unit}
\acro{apci}[APCI]{Application Protocol Control Information}
\acro{asdu}[ASDU]{Application Service Data Unit}
\acro{io}[IO]{information object}
\acroplural{io}[IOs]{information objects}
\acro{cot}[COT]{cause of transmission}
\acro{mitm}[MITM]{Man-in-the-Middle}
\acro{fdi}[FDI]{False Data Injection}
\acro{ids}[IDS]{intrusion detection system}
\acroplural{ids}[IDSs]{intrusion detection systems}
\acro{siem}[SIEM]{Security Information and Event Management}
\acro{mv}[MV]{medium voltage}
\acro{lv}[LV]{low voltage}
\acro{cdss}[CDSS]{controllable distribution secondary substation}
\acro{bss}[BSS]{battery storage system}
\acroplural{bss}[BSSs]{battery storage systems}
\acro{pv}[PV]{photovoltaic}
\acro{mp}[MP]{measuring point}
\acroplural{mp}[MPs]{measuring points}
\acro{dsc}[DSC]{Dummy SCADA Client}
\acro{fcli}[FCLI]{Fronius CL inverter}
\acro{fipi}[FIPI]{Fronius IG+ inverter}
\acro{sii}[SII]{Sunny Island inverter}
\acro{tls}[TLS]{Transport Layer Security}
\acro{actcon}[ActCon]{Activation Confirmation}
\acro{actterm}[ActTerm]{Activation Termination}
\acro{rtt}[RTT]{Round Trip Time}
\acro{c2}[C2]{Command and Control}
\acro{dst}[DST]{Dempster Shafer Theory}
\acro{ec}[EC]{Event Correlator}
\acro{sc}[SC]{Strategy Correlator}
\acro{ioc}[IoC]{Indicator of Compromise}
\acroplural{ioc}[IoCs]{Indicators of Compromise}
\acro{ot}[OT]{Operational Technology}
\acro{BSS}[BSS]{Battery Storage System}
\acro{cpt}[CPT]{Cyber-Physical Twin}
\acro{arp}[ARP]{Address Resolution Protocol}
\acro{soc}[SOC]{State of Charge}
\acro{ems}[EMS]{Energy Management System}
\acro{mac}[MAC]{media access control}
\acro{hitl}[HITL]{hardware-in-the-loop}
\acro{iot}[IoT]{Internet-of-Things}
\acro{dt}[DT]{Digital Twin}
\acroplural{dt}[DTs]{Digital Twins}
\acro{cps}[CPS]{cyber-physical system}
\acro{dsr}[DSR]{Demand Side Response}
\end{acronym}

%% file: chapter1.tex
\section{Introduction} \label{sec:introduction}
The increasing penetration of \ac{ict} in the power grid~\cite{tondel2018interdependencies} is leading to new opportunities for active grid operation, especially at the distribution grid levels that form the backbone for \ac{sg} operation~\cite{davarzani2021residential}.
However, this increasing interconnection of systems and actors is leading to the emergence of threats~\cite{mathas2020threat} and thus a new threat landscape composed of cyberattacks that threaten the stable and reliable state of the grid~\cite{krause2021cybersecurity}.
To adequately address these new threats, mitigation and countermeasures must be integrated as an essential part of the grid infrastructure, consisting not only of preventive but also reactive measures~\cite{van2020methods}.
One challenge in integrating countermeasures in the form of security concepts, especially new and active mitigation measures, is evaluating their effectiveness and their own impact on the grid~\cite{dan2012challenges}.
This requires testing and validation of these measures in these systems by replicating attack scenarios, which is often not feasible in productive operated grids~\cite{atalay2020digital}.
In addition, the development and validation of these measures, in particular data-driven approaches, requires attack data that often cannot be accessed~\cite{burstein2008toward}.
Alternatively, attack data could be generated synthetically, but approaches in this area are still under research and require further validation~\cite{abt2014plea}.
Another approach to this issue might be to generate attack data on the basis of replicas of the grid, such as \acp{dt}, that enable security investigations~\cite{alshammari2021cybersecurity}.

To provide a basis for studying the cybersecurity of the power grid by enabling the generation of attack data, a feasible approach is needed without compromising the grid itself.
The challenges we seek to address in this paper are therefore the following:
\begin{enumerate}[(i)]
    \item The replication of the grid including all layers of interest such as communication layer, process field and control layer in a safe and controlled environment.
    \item The execution of realistic attack scenarios without affecting the reference grid itself, thus avoiding harmful repercussions on the grid.
    \item The generation of attack data during the attack experiment, including process and communication data.
\end{enumerate}

Therefore, in this paper, we propose a \ac{cpt} approach that enables the investigation of power grid cybersecurity without compromising grid security.
This allows the generation of attack data through cyberattack replications in \ac{sg}.
To this end, we present an approach built on a co-simulation framework integrated into a laboratory environment with the ability to replicate \ac{sg} in its process and \ac{ict} layers, and demonstrate it using a real microgrid.
In particular, our contributions in this paper are:
\begin{enumerate}
	\item We present the current state of the art in the \ac{dt} based test and investigation environment and highlight the problem of attack data generation (Section~\ref{sec:background}).
	\item We describe the overall \ac{cpt} approach, which consists of the co-simulation framework, the laboratory environment, and the reference grid (Section~\ref{sec:framework}).
	\item We demonstrate and discuss the cybersecurity investigation capabilities of our proposed approach through a case study of cyberattack (Section~\ref{sec:result}).
\end{enumerate}

%% file: chapter2.tex
\section{Application of Digital Twins} \label{sec:background}
In this section, we present the state of the art in \ac{dt} approaches based on the use case of cybersecurity analysis.
In particular, Section~\ref{subsec:digital-twin} presents different research directions of \ac{dt} in \ac{sg}, while Section~\ref{subsec:background_detection} presents cybersecurity-oriented testbed approaches, and Section~\ref{subsec:background_problem} highlights current problems in attack data generation.

\subsection{Digital Twins in Smart Grid} \label{subsec:digital-twin}
In the context of this work, we understand the term \ac{dt} to describe a system that replicates a physical target system through continuous model-based simulation of the physical target system's functionalities~\cite{qian2022digital}.
This requires an exchange of data between the \ac{dt} and the physical target system, based on which the physical state of the system and its composition can be determined and updated.
The capabilities of the \ac{dt} can therefore be used to predict, control, and optimize the functionalities of the target physical system, while also interacting in the form of feedback to adapt to environmental changes.

Several use cases are being pursued with the \ac{dt} approaches, in particular the work of Olivares et al.~\cite{olivares2021towards} presents a \ac{dt} in advanced metering infrastructures within home area networks consisting of smart meters and \ac{iot}-devices.
Their approach includes a \ac{dt} framework controller that is responsible for orchestrating and coordinating physical state variables with the virtualized components that represent their physical counterparts.
The authors demonstrate their \ac{dt} approach by examining specific attack vectors to be detected by the \ac{dt} model.

In light of the increasing integration of \ac{iot}-devices and the associated lifecycle management challenges, the work by Atalay et al.~\cite{atalay2020digital} presents a \ac{dt}-based approach to \ac{sg} lifecycle security assessment.
Their approach includes threat modeling-based security assessment, where the \ac{dt} represents the target system and attack simulation tools are used in conjunction with risk assessment analysis to examine the level of security.
However, this work was presented only on a conceptual basis.

The work by et al.~\cite{kandasamy2021epictwin} presents a \ac{dt} designed with the goal of achieving high performance through the use of virtual machines and Docker containers.
They test the setup in real industrial control systems.
Furthermore, the twin replicates the target system's communication stack to enable seamless integration of existing security approaches, which use data-centric analysis methods, for benchmark testing.
In particular, the authors demonstrate their approach in a case study representing a zone consisting of multiple smart homes where a \ac{mitm} attack is carried out.

\subsection{Cybersecurity Analysis} \label{subsec:background_detection}
A particular use case that we are pursuing in this work is the investigation of cybersecurity of critical infrastructures such as power grids in a secure and controlled environment.
In this area, several research works have been conducted to investigate the security state of power grids using \ac{cps} testbeds.

The work of Oyewumi et al.~\cite{oyewumi2019isaac} presents a testbed for real-time experimental research aimed at investigating and monitoring the impact of potential cyberattacks and evaluating the performance of new cybersecurity solutions for \ac{sg}.
Based on \ac{scada} systems, the testbed uses real-time simulation of the power grid as well as remote human monitoring and control of substations and a TCP/IP stack for communication protocols.
The lab environment is designed to provide a secure and closed environment that facilitates forensic investigations, penetration testing, and experimental analysis to secure critical infrastructure.

The work by Noorizadeh et al.~\cite{noorizadeh2021cyber} presents a hybrid testbed for industrial control systems implemented by real-time simulation of the Tennessee-Eastman process as the physical component of the testbed is combined with emulation of field devices.
In their work, they address the challenge of obtaining hard-to-access field data by generating and logging the data from the physical part of the proposed testbed.
The authors emphasize the security investigation of the physical grid reference, especially based on the PROFINET protocol that was attacked by a \ac{mitm} scenario where real-time data injection was performed.

With respect to a combined environment consisting of a testbed and a co-simulation framework, the work by Hammad et al.~\cite{hammad2019implementation} presents an offline co-simulation approach for \ac{cps} power systems.
In particular, the authors pursue an offline co-simulation approach in which the time step of one or multiple simulation software is longer than that of the physical system being simulated to provide a more accessible \ac{sg} testbed.
The corresponding simulation components integrated using a discrete-event scheduler are the energy simulator PSCAD and the open-source communication simulator OMNeT++.

\subsection{Problem Analysis} \label{subsec:background_problem}
When analysing the cybersecurity of power grids, many problems and challenges must be considered.
In particular, the problem of missing and insufficient attack data in the critical infrastructure domain poses a challenge for the validation and testing of security concepts.
Many of the related works propose a \ac{dt} or a \ac{cps} testbed to replicate the target physical system and perform the security investigation without compromising the real grid reference.
However, some of these approaches integrate simulations, virtualized components, and simplified communication interaction models into their investigation environment, which can potentially affect the quality of the output data.
Considering a fully physical operated testbed is, however, not feasible due to the costly and inefficient scaling factor.
Moreover, it is often not flexible enough to fit diverse use case scenarios.
Therefore, a hybrid approach, where certain grid segments of interest are fully replicated in a physical environment while the remaining grid segments are virtualized, may provide a suitable trade-off between data quality and flexibility.
This could be, in particular, a flexible constellation of \ac{hitl} simulation of selected grid segments enabled by coupled simulators to simulate environmental effects by segments not considered in the physical laboratory.
Moreover, attack scenarios with real attack vectors can be fully emulated within the physically replicated segments, avoiding the use of simplifications and assumptions, thus improving overall data quality.
Consequently, in this paper, we take a hybrid approach between \ac{cps} testbed and \ac{dt} in the form of a \ac{cpt}, on which basis we perform security analysis within cyberattack case study fully emulated in the environment.

%% file: chapter3.tex
\section{Cyber-Physical Twin Approach} \label{sec:framework}
In this section we introduce our \ac{cpt} approach.
First, we provide in Section~\ref{subsec:framework_overview} a general overview about the approach and the architecture.
Second, we describe in Section~\ref{subsec:grid-reference} the real-world grid segment and the laboratory environment we use for the \ac{cpt} in Section~\ref{subsec:lab-environment}.
Finally, we present in Section~\ref{subsec:hil} the \ac{hitl} Simulation used for the control of our laboratory.

\subsection{Approach Overview} \label{subsec:framework_overview}
\begin{figure}
	\centering
	\includegraphics[width=\linewidth]{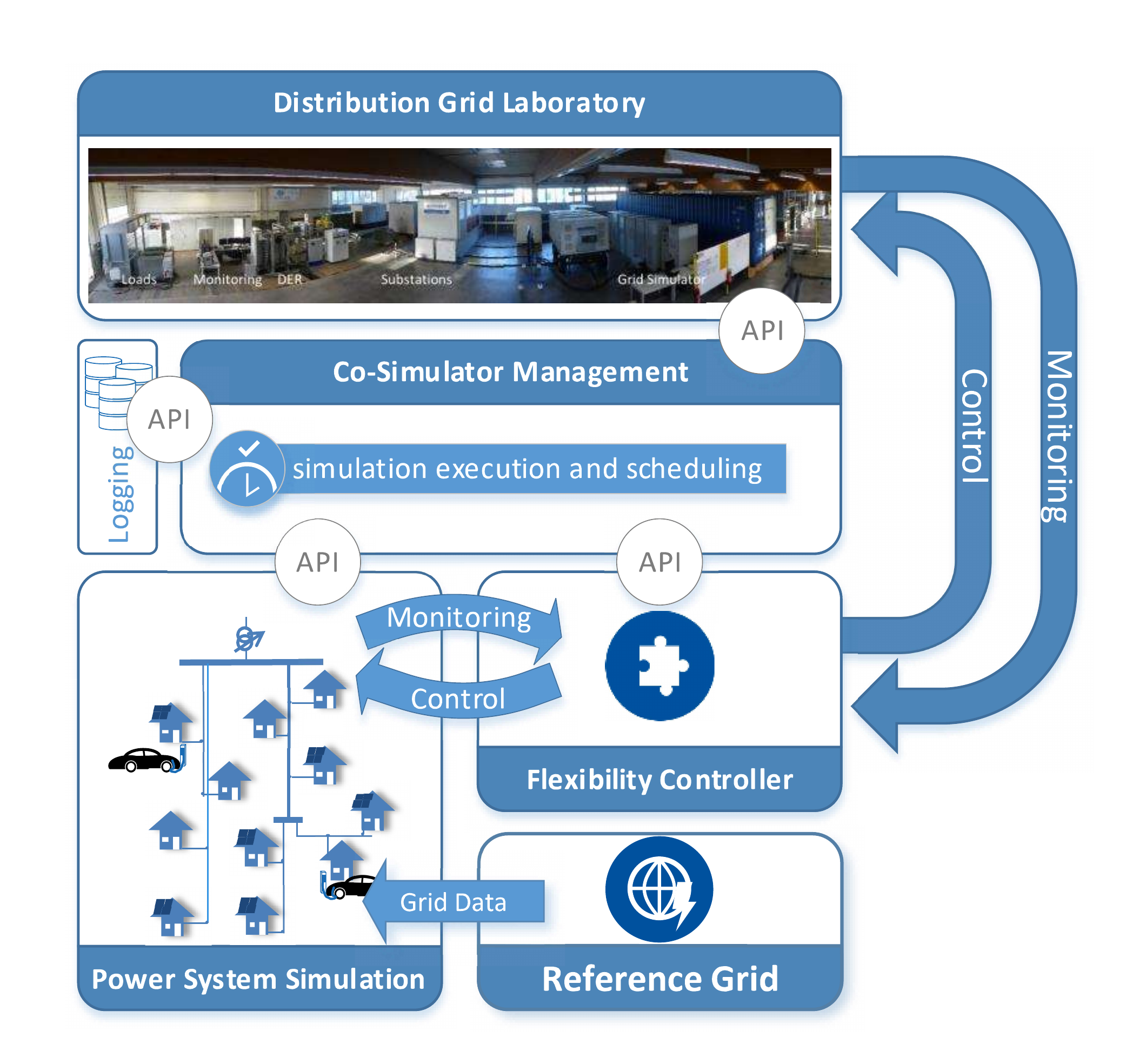}
	\caption{Overview of the \ac{cpt} approach consisting of the co-simulation framework and the \ac{sg} lab.}
	\label{fig:cpt_approach}
\end{figure}

Figure~\ref{fig:cpt_approach} illustrates the overview of our \ac{cpt} approach.
The \ac{cpt} in our laboratory environment is a replica of a real grid segment in Italy.
The key characteristics of the real-world topology and assets are replicated in the encapsulated environment.
Using multiple measurements of the grid segment, we can control loads and generators to emulate the real-world gird utilisation.
We implement management algorithms for controllable assets like \acp{bss} based on the residual load measured in the laboratory grid.
Therefore, we are able to test those algorithms independently of the real-world situation and without effecting the normal operation in Italy.
On this basis, we are also able to carry out cyberattacks on the \ac{ict}.
This allows us to carry out a realistic cyberattack without affecting normal operations in the respective grid segment.

\subsection{Grid Reference} \label{subsec:grid-reference}
The grid segment for which we are creating the \ac{cpt} is located in Terni, Italy, at the headquarters of the local multi-utility operator ASM.
The initial grid consists of two secondary substations with two sub-grids.
There is one cross branch connecting the underlying parts of the substations.
The grid segment consists of two large-scale \ac{pv} systems.
Besides the normal loads, two charging points for electric vehicles exist - one fast-charging-station and one with a power of 22~kW.
Also, one \acs{bss} exists in the grid segment.

There are seven measurement points to monitor characteristic values like power, voltage, and, frequency.
Each measurement device uploads the measurements via MQTT broker.
Therefore, our \ac{cpt} performs power flow analysis and management algorithm replication based on these measurements.

\subsection{Laboratory Environment} \label{subsec:lab-environment}
In this section, we describe the structure of our laboratory environment as the core of our \ac{cpt} approach.
Due to physical constraints, it is not feasible to replicate the exact structure of the grid segment in our lab.
Therefore, we must create a suitable grid structure, taking into account the available resources, first.

Our laboratory environment consists of multiple \ac{pv} inverters with maximum power in the range of 10~kW to 36~kW,
multiple loads with a maximum power of 20~kW and 45~kW and a \ac{bss} inverter with rated power of 15~kW.
In addition to those end-user devices, a secondary substation with a rated power of 630~kVA is also part of our lab.
Given the physical limitations of scalability, the measured values of the grid segment are scaled down.

\begin{figure}
	\centering
	\includegraphics[width=\linewidth]{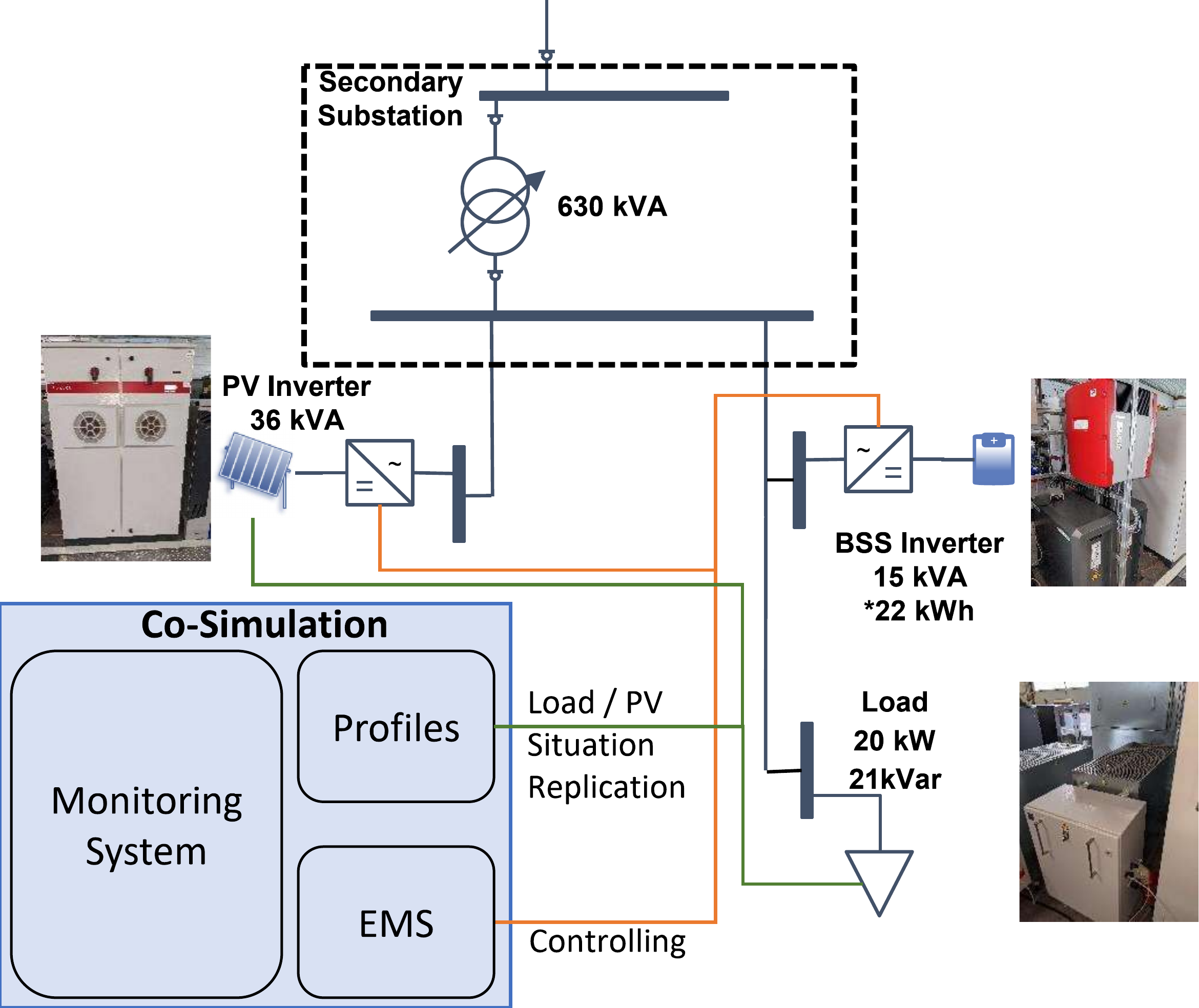}
	\caption{Grid topology of the \ac{cpt} in our \ac{sg} laboratory.}
	\label{fig:cpt_lab}
\end{figure}

We can create various grid topologies using multiple low-voltage cables and distribution boxes.
Figure~\ref{fig:cpt_lab} illustrates in a simplified view the topology we created in our laboratory environment.
There are two branches connected to the secondary substation.
The \ac{pv} inverter with a rated power of 36~kW is connected to the left branch.
To ensure controllable generation profiles, the inverters are connected to controllable DC power supplies.
The \ac{bss} inverter (15~kW / 22~kWh) and a load bank (20~kW) are connected to the right branch.
To monitor the power flow in our laboratory grid, we install \textit{Janitza} power analysers at the inverters grid crossing points and the low voltage level of the transformer.
In addition, we also install network taps to capture the communication traffic.

Besides the electrical connection, we also construct an \ac{ict} infrastructure.
We connect the end-user devices via their Ethernet interfaces over a replicated process network to an \ac{ems}.
It communicates with the assets using the Modbus TCP protocol. 
The \acs{ems} logic is embedded in the framework implemented for the \ac{hitl} simulation.

The \ac*{cpt} is not an exact duplicate of the real-world grid topology and\ac*{ict}.
Due to limited information about the detailed grid topology and limited asset resources in our laboratory we can replicate only a similar infrastructure.
Nevertheless, it is possible to scale down the system to appropriate power limits and replicate the measured power for generators and consumers.
Therefore, we can replicate the general behaviour of the real-world system by our \ac*{cpt} with sufficient accuracy for our case of investigation.

\subsection{\ac{hitl} Simulation} \label{subsec:hil}
For the time-dependent control of our laboratory, we implemented a \ac{hitl} simulation environment.
This environment is implemented as \textit{co-simulation} environment~\cite{van2021towards}.
It allows time-discrete simulation steps.
Part of the co-simulation environment are various simulators.
A central scheduler handles the connections between these simulators and manages the exchange of information at each time instance.
The advantage of the proposed approach is the modular design and the flexible scheduling process.
For this reason, simulated components and real world assets flexibly can be connected and analysed.
The depth of detail can vary for different investigation scenarios depending on evaluated part of the system.
E.g. the communication part can be simulated only with the relevant exchange of information or communication adapters can be used for protocol conform communication.

\begin{figure}
	\centering
	\includegraphics[width=\linewidth]{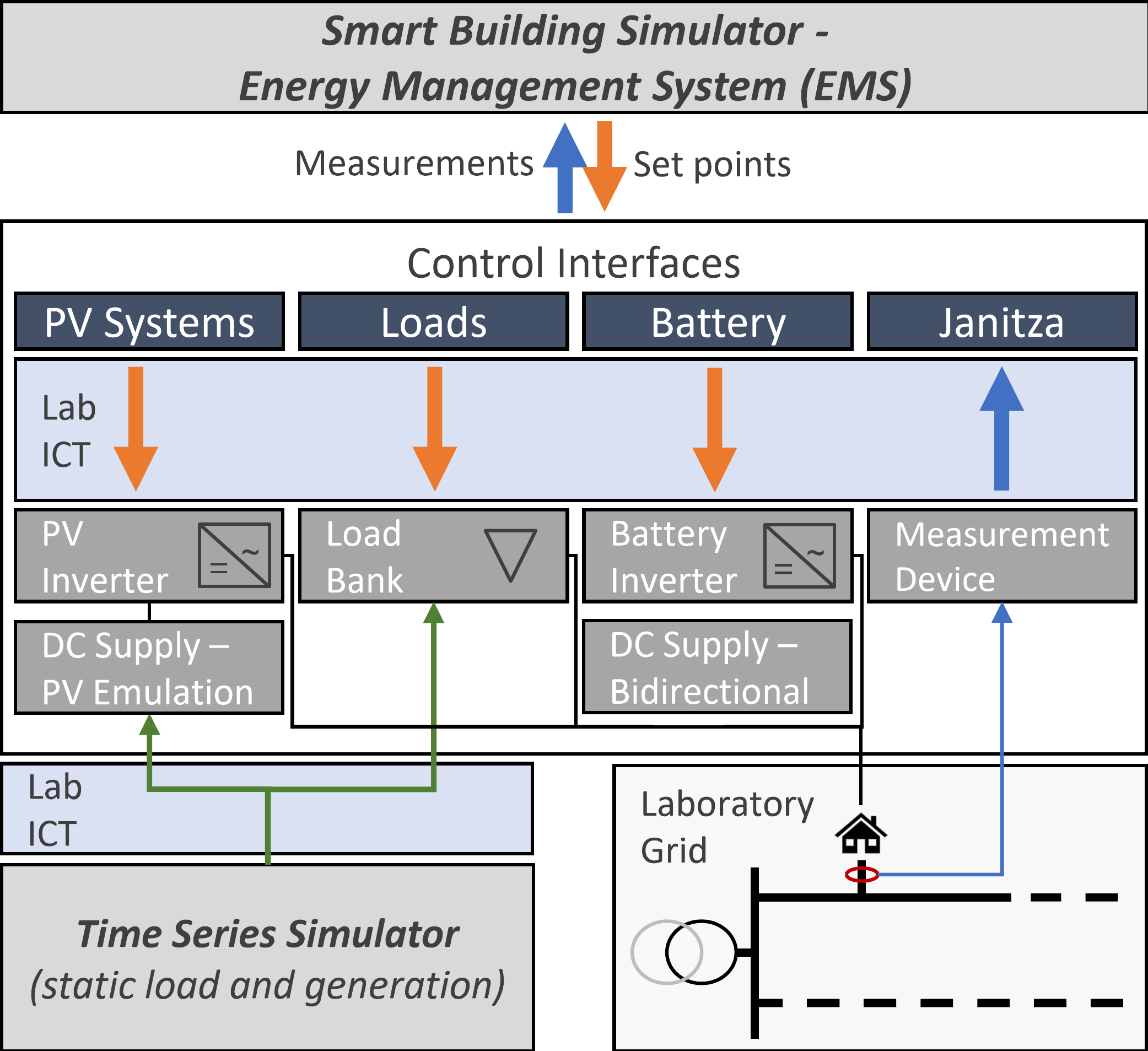}
	\caption{Co-Simulation - Schematic scheduling and controlling process. Black:~electric connection; Blue:~measurements; Orange:~Control commands; Green:~Profiles}
	\label{fig:co-sim}
\end{figure}

The \textit{co-simulation} framework allows connecting various simulators and organizes the time scheduling process and data management.
Figure~\ref{fig:co-sim} shows the schematic connection of simulated parts and controlled assets.
Relevant for this use case are the time-series simulator, the controller for the DC supply and the load bank, and the simulator representing the building management including the \acs{ems}.
The co-simulation is the connector for the management algorithm implemented as software module and the assets controlled in the laboratory.
Due to the investigation of attacks on the communication infrastructure, the protocol conform communications adapters are used.
The time-series simulator applies the measurement data from the real grid to the grid segment in the laboratory.
Those fixed time series for demand and generation are then used as input for the controller of the DC supply and the load bank.
The simulator for the building management system includes an \ac{ems} as logic part and communication and control interfaces for the assets in the grid segment.

In the normal operation mode the \ac{ems} manages the load balancing using controllable assets.
For this, real-time measurement data provided by the \textit{Janitza} measurement device are used.
Based on the measurement data, the \ac{bss} tries to shave load and solar peaks.
Therefore, the \ac{ems} processes the measurement data and sends control commands to the flexible assets.
In the real-time simulation measurement data is collected at each scheduled time instance.
The data processing including the determination of new control commands also happens in real time.

%% file: chapter4.tex
\section{Case Study \& Discussion} \label{sec:result}
In this section, we demonstrate our \ac{cpt} approach in the context of a selected case study and discuss the results with respect to its use case in the cybersecurity domain.
First, we introduce the cybersecurity condition adopted in the case study in section~\ref{subsec:security}.
Afterwards, we describe our case study scenarios in Section~\ref{subsec:result_proecdure}, while we present the case study results in Section~\ref{subsec:result_res} and discuss the results in Section~\ref{subsec:result_dis}.

\subsection{Cybersecurity Condition in Case Study} \label{subsec:security}
\begin{figure}
	\centering
	\includegraphics[width=\linewidth]{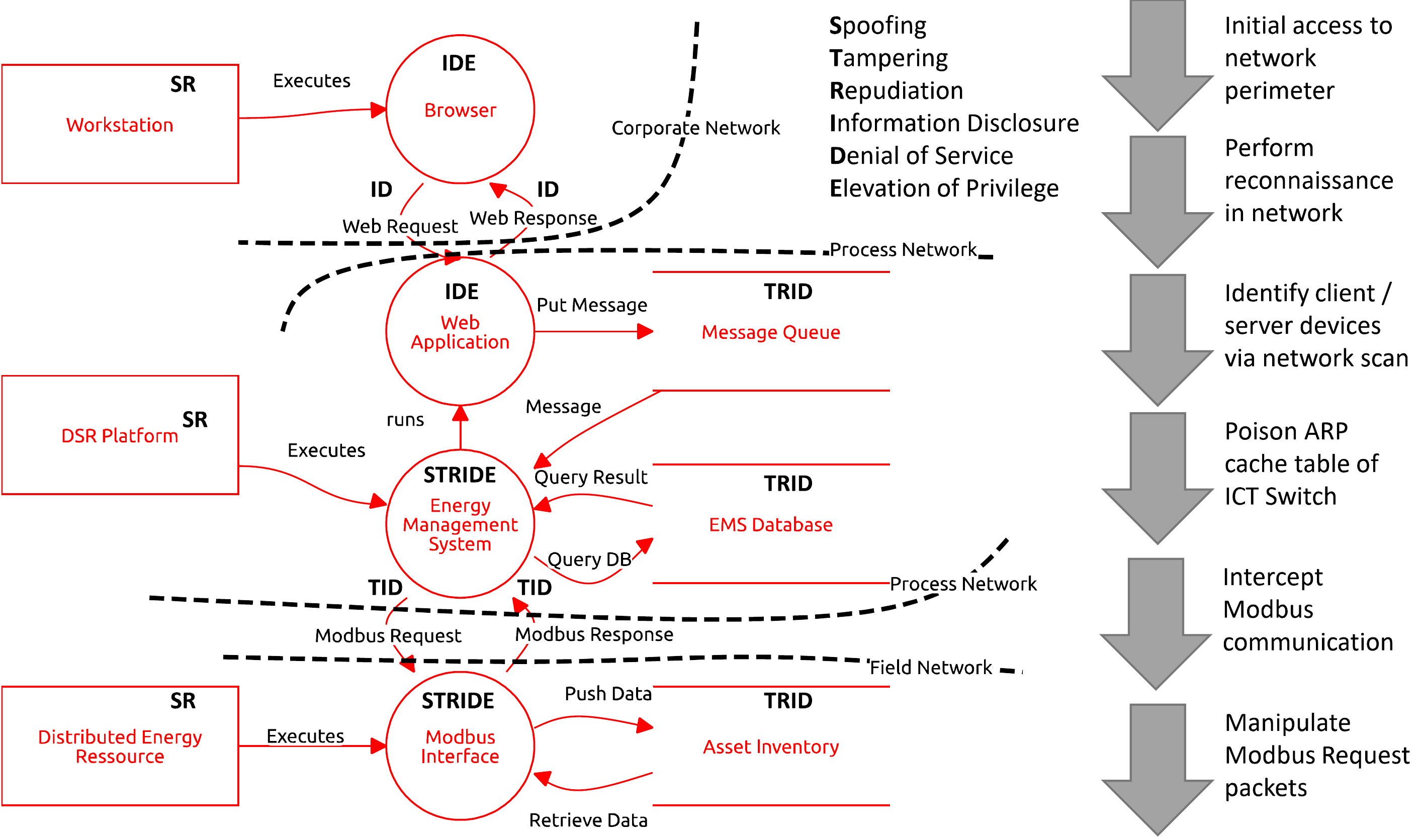}
	\caption{STRIDE-based threat model for the case study scenario.}
	\label{fig:results_threatmodel}
\end{figure}
To proceed with the study of cybersecurity using the \ac{cpt} approach, we first establish the security state conditions in our lab.
The relevant components used in the lab are the workstation as a \ac{dsr} platform, the load bank with integrated \ac{plc} for remote control, and the \ac{der} inverters connected via a \ac{ict} switch.

A common vulnerability within the site is the outdated state of software and services that are not patched with the latest security updates or are no longer supported due to end-of-life.
The \ac{dsr} platform, as the control system within the use case, runs on an outdated operating system with no security updates or support.
Specifically, the critical vulnerabilities within the component are insecure remote control functions (e.g., remote desktop protocol), a weak password hash function, and default credentials.

When examining other components within the system, default credentials that are sometimes difficult to change or not requested for critical operations are a major security concern.
This is especially a typical characteristic for the \ac{ot} facilities with default password that can be searched online.
In addition, communication channels within the network do not use security mechanisms such as encryption or authentication.
Critical access points via facility web interfaces are also not encrypted.
In particular, the use of legacy protocols such as Modbus lacks security mechanisms and often contains functions that can be accessed without authentication procedures that expose process information.

The resource-constrained conditions of the hardware also allow for potential attacks on the endpoint without adequate safeguards.
Table~\ref{tab:results_secfunctable} summarizes the security and functional features of the case study.
\begin{table}[]
\centering
\caption{Overview of security and functional features}
\resizebox{\columnwidth}{!}{%
\begin{tabular}{@{}lll@{}}
\toprule
Component &
  Function &
  Security Issue \\ \midrule
\textit{DSR Plattform} &
  Optimization of self-consumption via controlling of DER &
  \begin{tabular}[c]{@{}l@{}}-EOL of OS\\ -Legacy protocols\\ -Outdated software\\ -Weak password protection\end{tabular} \\
DER Inverter &
  IED functions for remote monitoring and control &
  \begin{tabular}[c]{@{}l@{}}-Legacy protocols\\ -Default password\\ -No encryption\end{tabular} \\
Load Bank &
  Replication of the load profiles of the reference system &
  \begin{tabular}[c]{@{}l@{}}-Legacy protocols\\ -Default password\\ -Fragile software\end{tabular} \\ \bottomrule
\end{tabular}%
}
\label{tab:results_secfunctable}
\end{table}
Overall, the security conditions in the lab environment reflect typical security conditions in \ac{ot} networks that have neglected security concerns of their systems due to historically isolated and closed environments that are now exposed over the Internet.

\subsection{Case Study Scenarios} \label{subsec:result_proecdure}
\begin{figure}
	\centering
	\includegraphics[width=\linewidth]{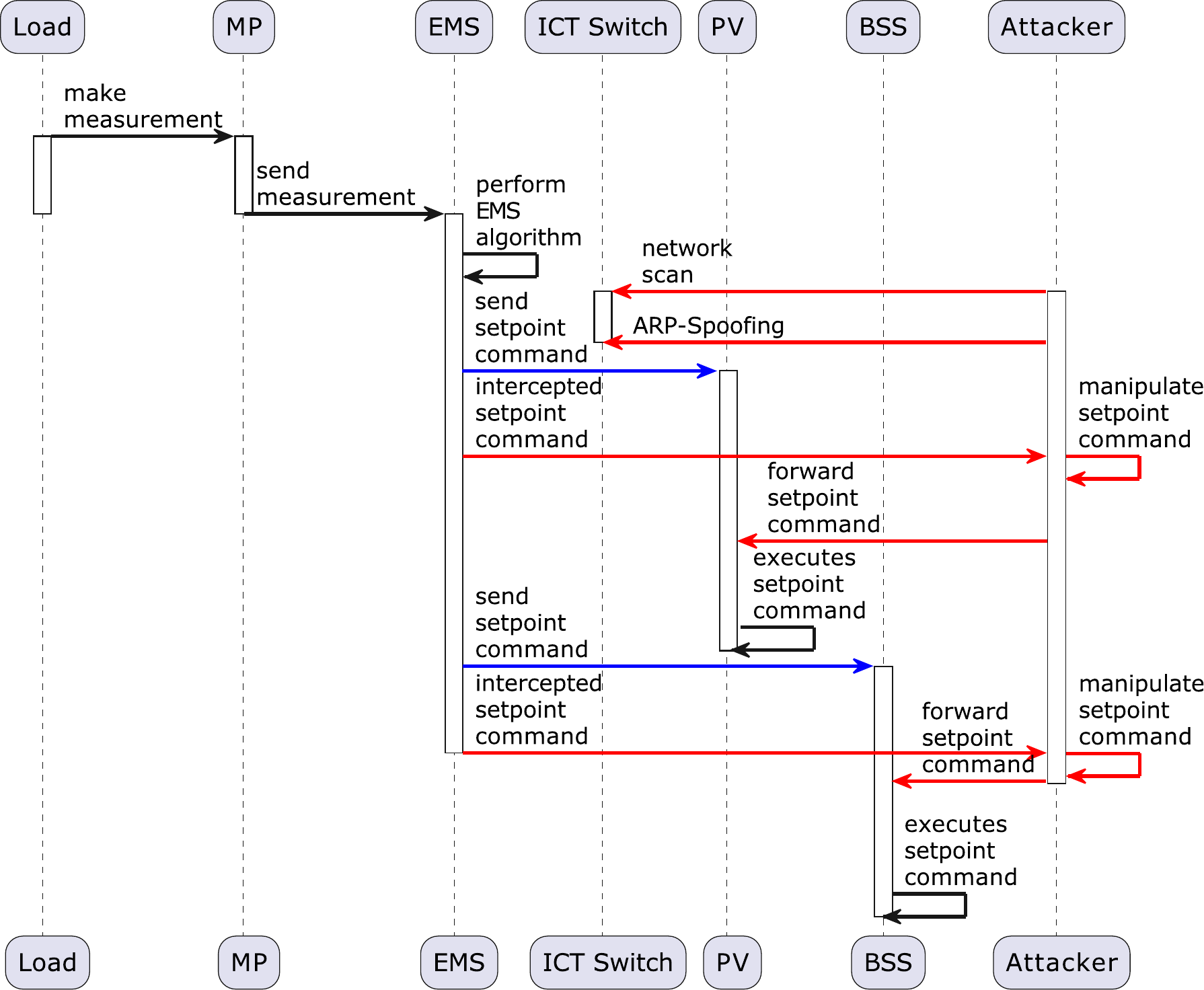}
	\caption{Sequence diagram of the case study scenarios. One scenario represents normal operation (indicated by blue arrows) and one represents the attack scenario (indicated by red arrows).}
	\label{fig:results_scenario}
\end{figure}
In our case study, we examine two scenarios for the \ac{cpt} approach (cf. Figure~\ref{fig:results_scenario}).
The first scenario is the reference scenario, which is the normal grid situation replicated based on the grid reference.
This scenario also serves as the benchmark against which the attack scenario is compared.
The second scenario is the attack scenario in which cyberattacks against the grid within the infrastructure with the objective to maximize the power imbalance are replicated
Based on the STRIDE threat model~\cite{shostack2008experiences}, the threats and attack vectors of the case study are identified (cf. Figure~\ref{fig:results_threatmodel}).
In particular, the attacker performs a sequence of attack actions composed of network scanning, role identification, \ac{mitm}, and control command manipulation to achieve its final objective.
Specifically, for the \ac{mitm} attack, we use \ac{arp} spoofing, which poisons the switch's \ac{arp} table to associate the target system's IP address with the attacker's \ac{mac} address to forward packets to the attacker instead of the original targets.
For instance, the attack scenario sets a power limit for the \ac{pv} inverter and the maximum charging power for the \ac{bss}.
The power limit for the \ac{pv} inverter is 3.5 kW and the maximum charging power for the \ac{bss} is 14 kW.
For the \ac{bss}, this means that it behaves like an additional load.
The simulation starts at 9:15am (real time) and the attack starts at 11:30am.
The duration of the attack is 1h 45 min and thus ends at 2:15 pm.

\subsection{Case Study Results} \label{subsec:result_res}
\begin{figure*}
	\centering
	\includegraphics[width=0.8\textwidth]{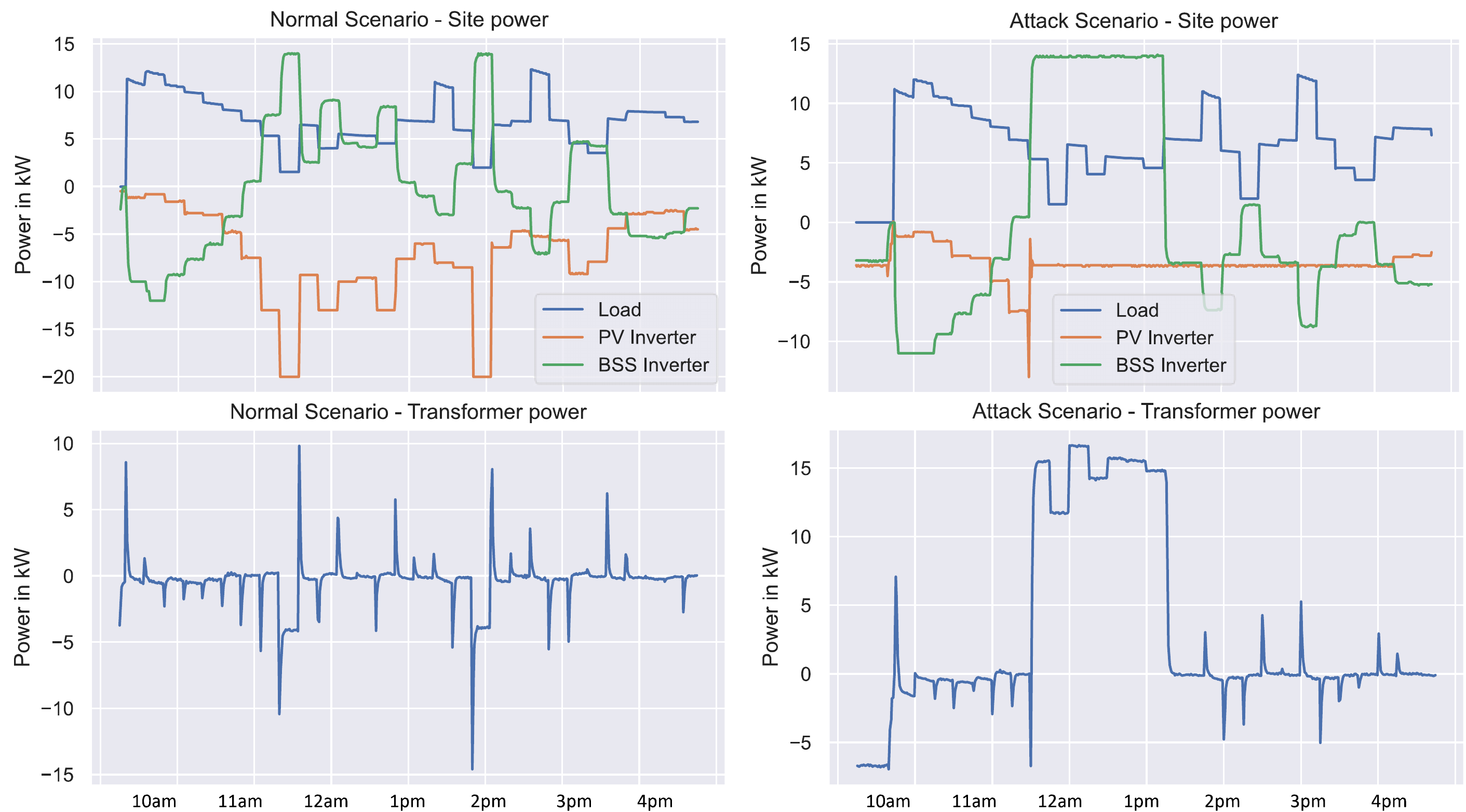}
	\caption{Results of the normal scenario (left) and attack scenario (right) which is conducted between 11:30 am and 2:15 pm.}
	\label{fig:results_plot}
\end{figure*}

\begin{figure*}
	\centering
	\includegraphics[width=0.8\textwidth]{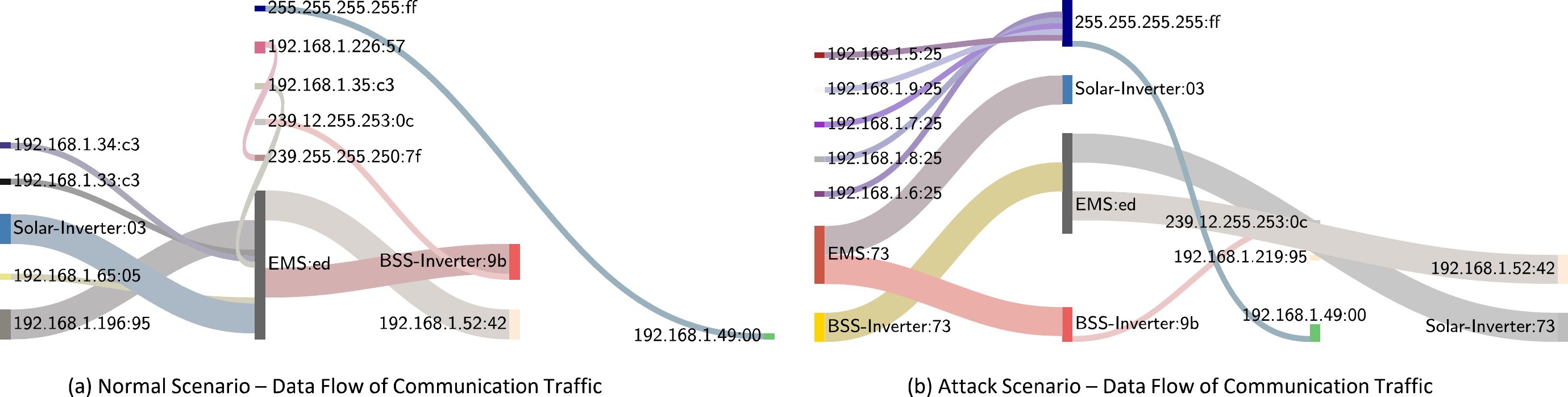}
	\caption{Data flow of the normal scenario (a) and attack scenario (b) which is conducted between 11:30 am and 2:15 pm.}
	\label{fig:results_plot_2}
\end{figure*}

Figure~\ref{fig:results_plot} illustrates on the left side the resulting power over time for the normal grid operation scenario.
Furthermore, Figure~\ref{fig:results_plot_2} additionally illustrates the data flow within the normal and attack scenarios of the experiment.
The load situation here is provided by the gird reference, where one day's data is used to replicate the load situation.
The high load situation within this reference is very early in the day around 10 am as well as spread out over time throughout the day (e.g. 1:30 pm, 2:30 pm).
Based on this load situation, the \ac{ems} in our co-simulation performs autonomous coordination of \acp{der} to minimize power exchange with the external (overlaying) grid.
We can observe that the \ac{bss} tries to balance the residual power at any time.
When the \ac{pv} generation is low, the \ac{bss} supplies almost all the consumed power.
When the \ac{pv} generation is higher than the demand, the remaining power is used to recharge the \ac{bss}.
The transformer output exhibits some ripple due to the short time required for the power management algorithm to rebalance the output.
Regarding the communication structure within the \ac{cpt} approach, it can be seen that the \ac{ems} communicates bidirectionally with the controllable devices such as \ac{pv} inverter and \ac{bss} with the star-shaped communication flow.
This indicates the centralized communication structure of the control system based on a web technology-based \ac{dsr} platform that could be deployed in a cloud infrastructure.

When looking at the attack scenario, certain characteristics can be identified.
In Figure~\ref{fig:results_plot} on the right side, it can be observed that the attack (11:30 am - 2:15 pm) increases the \ac{bss} power from one moment to another, while the \ac{pv} power decreases.
Therefore, the total power during the attack period is constantly very high.
Compared to the normal scenario, this is a significant difference.
Normally, the power should be balanced using the \ac{bss}, which means that the total power at the transformer should be close to 0 kW.
We can observe that the \ac{bss} management works properly after the end of the attack scenario.
In contrast, the limitation set by the attack is not reset.
The \ac{pv} generation does not exceed the power limit.
Only when the power generation is below the power limit the power injected into the site grid changes.
This could be due to an error in the management algorithm, which does not actively set a power limit for times when there is no limit.
Looking again at the data flow structure of the communication traces within the attack scenario, it is noticeable that the star-shaped structure of the \ac{ems} is replicated with the \ac{arp} spoofing.
The real \ac{ems} communicates with the \ac{pv} and \ac{bss} inverters;
however, these are spoofed endpoints of the attacker, i.e., \ac{arp} spoofed the communication channel between \ac{ems} and the respective endpoints.
As can also be observed, the attacker's fake \ac{ems} communicates with the real inverter endpoints so that the communication channel is intercepted.
The representation of the data flows here implies a parallel structure set up by the attacker, where the original \ac{ems} is supposedly still communicating with the inverter endpoints, but when looking at the \ac{mac} addresses, it becomes clear that these are interfaces spoofed by the attacker.
However, the original inverter appears to be communicating with the \ac{ems}, but looking again at the \ac{mac} addresses shows that the \ac{ems} interface has also been spoofed by the attacker.
This creates what appears to be a parallel data flow structure, with the original \ac{ems} communicating with the attacker's spoofed interfaces, while the inverters communicate with the \ac{ems}'s spoofed interfaces.

\subsection{Discussion} \label{subsec:result_dis}
In our case study, the scenarios replicated in the \ac{cpt} environment demonstrate the capabilities of a \ac{hitl}-oriented simulation environment that not only provides a basis for investigating security conditions in the real productive operated grids without actively intervening the grid operations.
A key vulnerability exploited by the cyberattack is the legacy characteristics of the technology, particularly the industrial protocols that do not provide for security mechanisms such as encryption or authorization.
The network architecture also contributed to the success of the attack, as it has a flat structure with no enforced segmentation or access control.
A vulnerable part of the system is also the \ac{dsr} platform, which runs outdated software that provides an attack surface for direct interference with the control logic.
Since the \ac{cpt} environment can only replicate a specific grid segment due to the physical constraints of scalability of real hardware, the data generation of the experiment must focus on the specific grid segments that are emulated in the \ac{cpt}.
However, the co-simulation component of the \ac{cpt} environment can be used as a substitute for the missing hardware to replicate the entire grid by virtually extending the replicated scenario.
Thus, a hybrid approach based on \ac{hitl} and \ac{dt} simulation can be pursued, where the \ac{hitl} simulation is used to replicate a specific grid segment in the physical environment and the \ac{dt} simulation is used to fully replicate the remaining segments in the digital world.
Thus, the \ac{cpt} environment can also be used as a validation reference for simulation.
Nonetheless, the \ac{cpt} approach in this paper demonstrates the capabilities of investigating specific coordinated cyberattack scenarios within a combined physical and virtual simulated environment that can be used for detailed security investigation of the grid reference without affecting the real grid.

%% file: chapter5.tex
\section{Conclusion} \label{sec:conclusion}
The increasing digitization of the power grid will lead to an emerging threat landscape for which countermeasures must be developed, but are faced with limited data from critical infrastructure during an attack.
In this paper, we demonstrate that our \ac{cpt} approach provides a basis for replicating severe cyber incidents in a secure, controllable, and isolated environment that includes a realistic grid and \ac{ict} infrastructure with real-world components.
In our case study, we observed the impact of specifically replicated cyberattacks, such as control command interception in a \ac{sg} use case where \ac{der} components are provided with control commands that are harmful to the grid.
Using advanced measurement infrastructure, we collected process and communication data during the cyberattacks.
In particular, this approach can provide dataset to support the development, testing, and validation of such detection approaches that leverage not only the data within the communication layer but also inconsistencies or anomalies within the process data.
Future work will therefore include the creation of a benchmark dataset that replicates various attack scenarios in the grid with coordinated attack patterns that can be used to improve countermeasures.